\documentclass[final,5p,times,twocolumn]{elsarticle}
\usepackage{amssymb}
\usepackage{subfigure}
\usepackage{courier}
\usepackage{xcolor}
\usepackage{colortbl} 

\usepackage[strict]{changepage}

\usepackage{framed}

\definecolor{formalshade}{rgb}{0.95,0.95,1}
\definecolor{darkblue}{rgb}{0.19, 0.55, 0.91}
\definecolor{lightgrey}{gray}{0.85}

\definecolor{pinkx}{rgb}{0.9,0.1,0.9}

\journal{}

\begin{document}

\begin{frontmatter}




\title{Can ChatGPT advance software testing intelligence? {An experience report on} metamorphic testing}

\author[a,b]{Quang-Hung Luu}
\author[a]{Huai Liu}
\author[a]{and Tsong Yueh Chen}

\affiliation[a]{organization={Department of Computing Technologies, Swinburne University of Technology},
    addressline={1 John Street, Hawthorn}, 
    city={Melbourne},
    postcode={3122}, 
    state={Victoria},
    country={Australia}}
\affiliation[b]{organization={Department of Civil Engineering, Monash University},
    addressline={14 Alliance Lane, Clayton}, 
    city={Melbourne},
    postcode={3168}, 
    state={Victoria},
    country={Australia}}

\begin{abstract}
While ChatGPT is a well-known artificial intelligence chatbot being used to answer human's questions, one may want to discover its potential in advancing software testing. 
We examine the capability of ChatGPT in advancing the intelligence of software testing {through a case study} on metamorphic testing {(MT)}, a state-of-the-art software testing 
{technique}. 
We {ask} ChatGPT to generate {candidates of} metamorphic relations (MRs), which are basically necessary properties of the {object program and which traditionally require} human intelligence to identify. These {MR candidates} are then evaluated in terms of correctness by domain experts.
We show {that ChatGPT} can be used to generate new {correct MRs} to test several software systems. Having said that, the majority of {MR candidates} are either defined vaguely or incorrect, especially for {systems that {have} never been tested with {MT}}. 
ChatGPT can be used to advance software testing intelligence by {proposing} MR candidates that can be later adopted for implementing tests; but human intelligence should still {inevitably} be involved to justify and {rectify their} correctness.
\end{abstract}


\end{frontmatter}


\section{Introduction}

ChatGPT, an advanced large language model (LLM) system developed by OpenAI, has recently {attracted global attention thanks} to its extensive expertise and remarkable conversational abilities. 
{It has} demonstrated proficiency in solving mathematical and logical problems \cite{frieder2023arxiv} as well as aiding bioinformatics research \cite{sima2023arxiv}. {In the} domain of software engineering, ChatGPT possesses remarkable capabilities {not only in} generating targeted source code {but also in generating} test cases and {assisting} debugging {programs \cite{celerdata2023}. {Nevertheless}, a recent {report} \cite{borji2023arxiv} has pointed out several shortcomings and failures of ChatGPT. Given the benefit of the doubt}, it remains uncertain whether ChatGPT can accomplish more sophisticated software engineering tasks, which typically involve complexity and necessitate a high degree of ``intelligence'', thereby posing greater {challenges}.

In this paper, we examine the potential of ChatGPT for enhancing software testing, specifically focusing on \textit{metamorphic testing} (MT), a leading-edge technique {endorsed by ISO/IEC/I-EEE \cite{iso2021mt}}. MT has been effectively utilized to identify faults in a wide variety of complex software systems, {including} Google Maps, Amazon, and {Facebook \cite{zhou2018,ahlgren2021}}. The remarkable success of MT can be attributed to its exceptional capability in tackling the test oracle problem, {a fundamental challenge} that has proven intractable for conventional software testing methodologies \cite{chen2017}. Furthermore, {MT} has gained recognition {as the most popular testing technique for} artificial intelligence (AI) and machine learning {(ML)} systems \cite{martinez2022}.

\begin{table*}[!htb]
\caption{Target systems for testing with ChatGPT.}
\label{table:programs}
\centering
\resizebox{18cm}{!}{%
\begin{tabular}{lllllccc}
\hline
ID & System & Description & Main inputs & Main outputs & Category & New system & No. MRs\\
\hline
1 & \textsc{sin}
    & program computing {\it sin} 
    & {one number} 
    & {one number} 
    & basic
    & -
    & 100 \\
2 & \textsc{sum}
    & program computing {\it sum} 
    & {a list of numbers} 
    & {one number} 
    & basic
    & -
    & 100 \\
3 & \textsc{shortest-path}
    & program finding the shortest path 
    & a graph with vertices, edges
    & {a path}
    & basic
    & -
    & 100 \\
4 & \textsc{regression}
    & multiple linear regression
    & multiple data row
    & coefficients, predicted data
    & non-AI/ML
    & -
    & 100 \\
5 & \textsc{fft}
    & {Fast Fourier Transform-based analysis}
    & time-series of data
    & frequencies, amplitudes
    & non-AI/ML
    & x
    & 100 \\
6 & \textsc{wfs}
    & weather forecasting system
    & multiple sources
    & multiple outputs 
    & non-AI/ML
    & x
    & 100 \\
7 & \textsc{av-perception}
    & autonomous vehicle perception 
    & images, point clouds
    & object detection
    & AI/ML
    & -
    & 100 \\
8 & \textsc{aicam}
    & AI-based home security camera 
    & images
    & object detection
    & AI/ML
    & x
    & 100 \\    
9 & \textsc{trafficsys}
    & AI-based traffic light control
    & sensor data
    & traffic decision
    & AI/ML
    & x
    & 100 \\
\hline
\end{tabular}
}
\end{table*}

MT makes use of the relations among multiple inputs and corresponding executions of the {target} program, being referred to as \textit{metamorphic relations} (MRs), to {check the correctness} of its implementation. If an MR is violated, the program is known to be faulty. For example, in testing the program $P$ to compute $\cos(x)$ ({where} $x$ is the angle in degrees), {it is difficult to judge the correctness of $P$'s output given any input (say, when $x = 12.34^\circ$), and thus to determine the pass/fail of $P$}. 
Given the {MR} $\cos(x+360^\circ)=\cos(x)$, we can generate the follow-up test case {$372.34^\circ$ from its source test case $12.34^\circ$}, and then validate the satisfaction of the MR, {that is, whether $P(372.34^\circ)=P(12.34^\circ)$ holds}. If the MR is violated, we know that $P$ is faulty.

The identification of new MRs is one of the greatest challenges in metamorphic {testing}, and thus has become a significant research focus in MT, which has attracted{, in our {recent search}, {over 60} publications}. 
{However}, current techniques for generating MRs are normally specific to certain application domains and rely on human intelligence. As the scope
of MT’s application continues its broad expansion, integrating
intelligent tools like ChatGPT may offer a significant potential to
advance this important research area.

With the emerging role of ChatGPT {in software engineering, it was shown to be able to help generate new MRs \cite{zhang2023chatgpt}}.  
{Instead of testing the ability of ChatGPT for {a specific system \cite{zhang2023chatgpt}}, this} study is targeted at {a wider range of systems and} seeking insights into the following questions{: (1) Can} ChatGPT generate correct MRs?
{(2) Can} ChatGPT generate {MRs for systems that have never been tested with MT}?
{(3) What} is the role of ChatGPT in {advancing software testing intelligence for MT}?

\vspace{-0.2cm}

\section{Method}

In this study, we {examined} the {effectiveness of ChatGPT (GPT-4)} in generating MRs. To assess its capabilities, 
{we chose nine target systems, as summarized in Table~\ref{table:programs}. Five of these systems have been tested with MT, while the other four have never been tested with MT, which are referred to as ``new target systems'' or simply ``new systems'' in the remaining of the paper.}
We classified these target systems into three categories, namely, basic, non-AI/ML, and AI/ML systems, to account for different complexity levels of software under test.

\begin{figure*}[!htb]
\centering
\subfigure[\it Clarifyng ChatGPT's understanding about the target system.]{\includegraphics[width=7.5cm]
{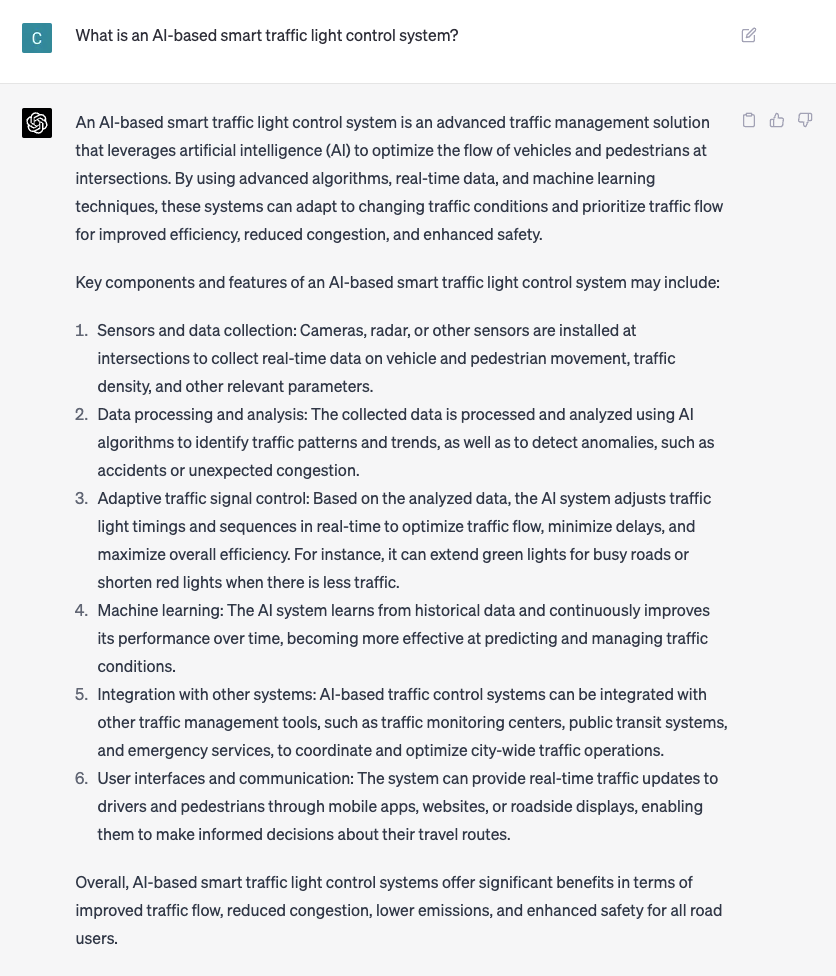}}%
\phantom{S}
\subfigure[\it Asking ChatGPT to generate MR candidates.]{\includegraphics[width=6.6cm]{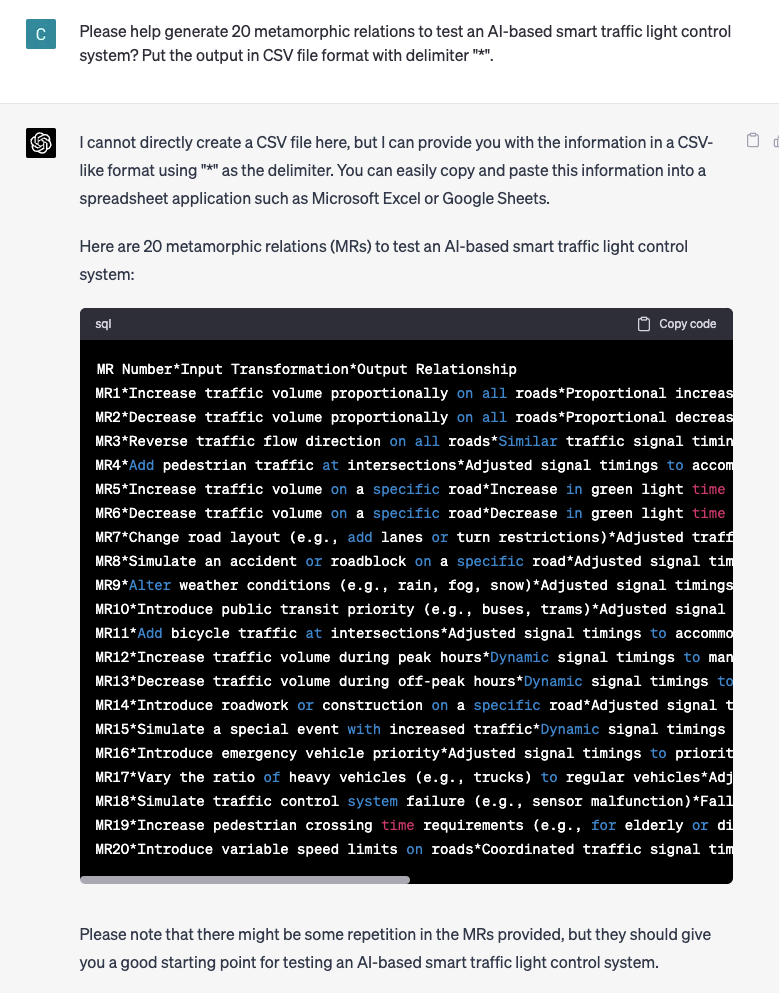}}%
\hfill
\caption{{Sample generation of new MRs with ChatGPT.}}
\label{fig:chatgpt:method}
\end{figure*}

{{To gain fair and useful answers} from what ChatGPT has been trained for, we used zero-short prompting in two steps (Fig.~\ref{fig:chatgpt:method}): first clarifying with ChatGPT about the target system, and then asking it} to generate a set of unique {MR candidates} for each target system {in each independent session. To overcome the limitation in answer's length of ChatGPT, we generated 20 MRs each time}. The correctness of {each candidate was} then evaluated by individual experts who have several years of experiences in both MT and the domain knowledge associated with the {systems. These candidates} are classified into three types, that is, {\it correct MRs} that were correctly and precisely defined; {\it incorrect MR candidates} that were incorrectly defined; and {\it unjustifiable MR candidates} that either (i) were imprecisely defined, or (ii) were vague and whose correctness cannot be judged.

{During our evaluation process, we noticed that when the number of MR candidates was large, their descriptions became repetitive.}  
Hence, we {focused} on examining the first 100 {unique MR candidates} for {each system}. {Both experts have {substantial agreement} with Cohen’s kappa coefficient of 0.86 and only a small discrepancy} in judging the MR candidates due to {the difference} in interpreting {their descriptions}, specifically, 64 out of 900 (around 7.1\%).
This discrepancy, mainly associated with understanding unjustifiable candidates, was then discussed and resolved {between two experts}.

\vspace{-0.3cm}

\section{Findings}

\begin{table*}[!htb]
\caption{Sample MR candidates generated by ChatGPT {and our revision} for {\textsc{regression}.}}
\label{table:results}
\centering
\resizebox{17cm}{!}{%
\begin{tabular}{p{1.6cm}p{14cm}p{2.4cm}p{1.5cm}}
\hline
Example & Description of MR candidate &  Source & Correctness\\
\hline
{MRC1} &
    Reorder data points, regression coefficients unchanged.
    & ChatGPT & Correct\\
\arrayrulecolor{lightgrey}\hline
{MRC2} &
    Changing the method used to estimate the regression coefficients, beta coefficients may be affected. 
    &  ChatGPT & Incorrect\\
\arrayrulecolor{lightgrey}\hline
{MRC3} &
    Replacing missing data with a constant value, beta coefficients may be affected.
    & ChatGPT & Unjustifiable\\
\arrayrulecolor{lightgrey}\hline
{MRC3-rev} &
    Replacing missing data with a constant value {\it on the regression line}, beta coefficients {\it are not} affected.
    & ChatGPT+Human & Correct\\
\arrayrulecolor{black}\hline
\end{tabular}
}
\end{table*}

\begin{figure*}[!htb]
\centering
\includegraphics[scale=0.52]{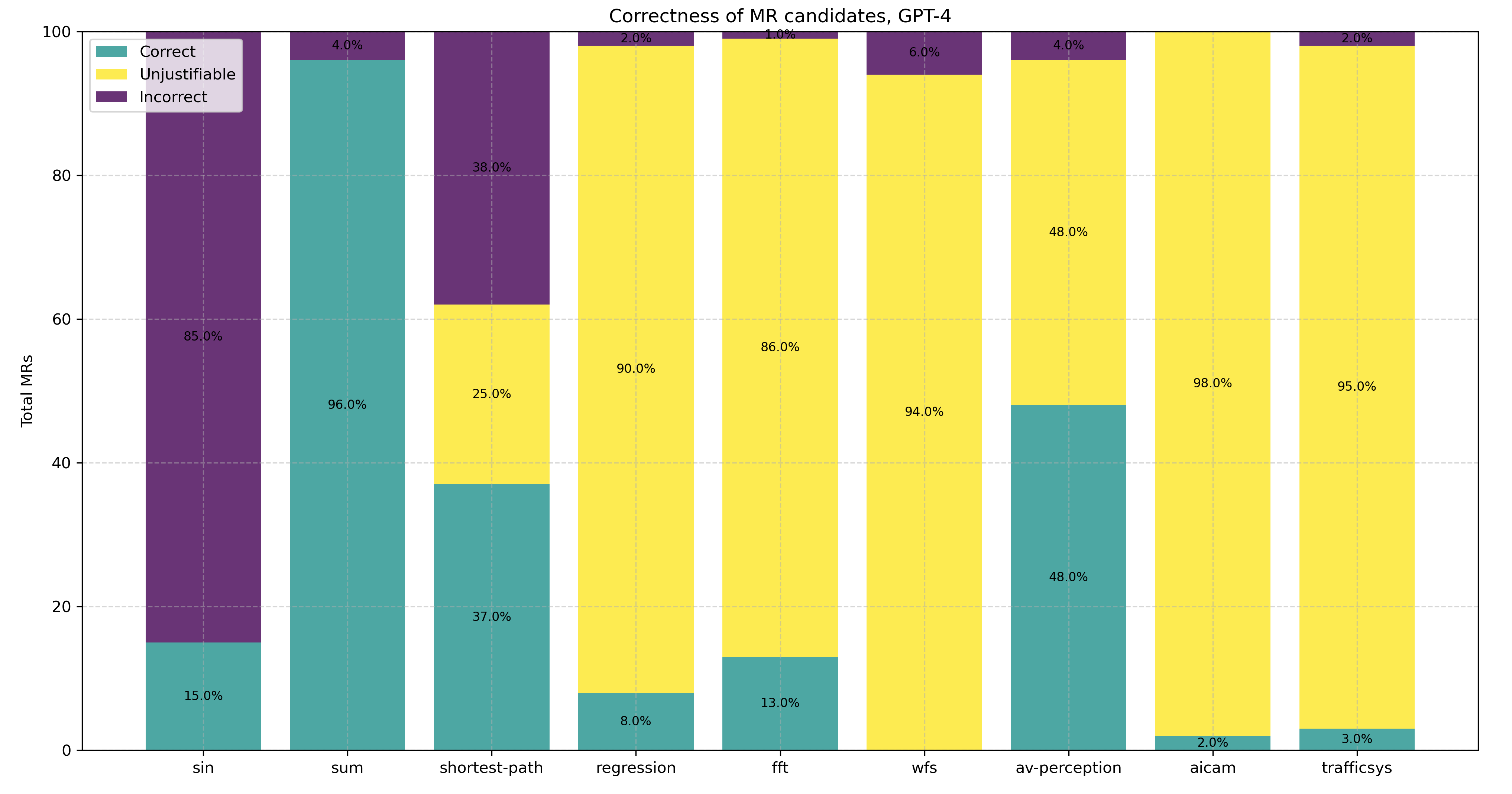}
\caption{{Evaluation of MR candidates generated by {ChatGPT}.}}
\label{fig:result1}
\end{figure*}

\subsection{Generation of correct MRs}

Table~\ref{table:results} presented some examples of {correct (MRC1), incorrect (MRC2) {and unjustifiable} (MRC3) MR candidates for the \textsc{regression}. MRC1} is a correct MR candidate because {the results of multiple linear regression processes do not depend on the order of data}. MRC2 is an incorrect MR candidate because {changing the method of regression is basically about using a different program under test. {Furthermore, it lacks a} description about the construction of follow-up input, leaving the definition of MR to be incorrect}. 
MRC3 is unjustifiable because the phrase ``may not be affected'' {means that the follow-up output may or may not change, which is always true. As a result, MRC3 will always be satisfied irrespective whether the program is faulty or not}. 

Overall, it {was} found that ChatGPT {could} generate correct {MR candidates} for almost all target systems/programs considered (Fig.~\ref{fig:result1}). On an average, ChatGPT {was} able to generate $49.3\%(=(15.0\%+96.0\%+37.0\%)/3)$ correct {MR candidates} for programs in the basic category {(Table~\ref{table:programs})}. For the programs that belong to {non-AI/ML} category, the arithmetic mean {was} {$7.0\%~(=(8.0\%+13.0\%+0.0\%)/3)$}. The mean value {was} slightly higher for the {AI/ML} category, which {was} {$17.7\%~(=(48.0\%+2.0\%+3.0\%)/3)$. 
{The} number of correctly generated {MR candidates} {varied} from {program to program}, with more correct {candidates} found in simple programs. 


\subsection{Generation of {MRs for new target systems}} 

{While ChatGPT might generate new MRs for systems that have been tested with MT, this study further showed that it could help propose MR candidates for new target systems that have never been tested using MT}. It could propose correct MR candidates for three out of these four target systems, namely, \textsc{fft}, \textsc{aicam} and \textsc{trafficsys} (Table~\ref{table:programs})}. 

We further {asked} ChatGPT to find and confirm the existence of research works that are relevant to these systems in the literature {for which ChatGPT returned} a list of 12 papers in the field of MT. 
We have further conducted a {thorough} search in {four popular databases (namely,} Elsevier's Science Direct, IEEE Xplorer, ACM Digital Library, Taylor \& Francis) and {additional query of 50 first pages with} Google Scholar. {{After these investigations}, we confirmed these systems have not been tested with MT before.} In a word, we argue that ChatGPT {had an intellectual capability in software testing, especially in {MT} because it can generate {MRs for new target systems}.}

\subsection{Role of ChatGPT in advancing software testing intelligence {for MT}} 

Despite various {successes}, ChatGPT {has} proposed many {MR candidates} which turned out to be incorrect. 
Surprisingly, 85\% of the MR candidates identified by ChatGPT for the extensively-studied sine program were incorrect.  
For the other systems, the percentages of {MR candidates} that could be judged either as correct or incorrect, {range} from 25\% to 97\%, indicating {a large} amount of unjustifiable candidates. Only for the programs that {involve} very basic mathematical knowledge (that is, \textsc{sin} and \textsc{sum}), could ChatGPT generate MR candidates that can be clearly classified as correct {or} incorrect.

We also noticed that several {MR candidates}, despite being {unjustifiable, do provide hints leading to the discovery of correct MRs. Revising the description of {{MRC3}} candidate} to be the one similar to {MRC3-rev (last row of Table~\ref{table:results})}  {would} address {the vague issues of MRC3}, and thus help construct a new correct MR from {an unjustifiable one}. 

These results {had} several implications for the adoption of ChatGPT. First, ChatGPT {might} make mistakes in generating {MR candidates, which {might} cause serious concerns if {they} {were} {directly} used to verify the quality of software systems. Secondly, not all {unjustifiable MR candidates} are {totally useless because they may provide insights leading to the discovery of new MRs}. {Last but not least}, human intelligence {was} still required in {rectifying and assuring} the correctness of these {MR candidates}.

\section{Conclusion}

{Our study} {suggested} that ChatGPT is an useful {tool {which} can handle testing tasks that require a certain level of intelligence}. 
In particular, it {could} be used to suggest innovative {MR candidates} for assessing {systems that {have} never been tested with MT}. Having said that, the definition of a large portion of these {MR candidates} {remained} {unjustifiable} or {incorrect}. In a word, while ChatGPT {might} provide valuable insights for formulating new {MR candidates}, human intelligence {remained} indispensable in {rectifying} their accuracy and appropriateness. 

{These remarks align well with the conclusion from a recent study using ChatGPT to generate MRs for autonomous driving systems    \cite{zhang2023chatgpt}.  Future work would further consider other behaviors of ChatGPT, such as its typical mistakes in generating MRs and systematic ways to convert unjustifiable MR candidates into correct MRs.}




\begin{thebibliography}{00}

\bibitem{frieder2023arxiv}Frieder, S., Pinchetti, L., Griffiths, R., Salvatori, T., Lukasiewicz, T., Petersen, P., Chevalier, A. \& Berner, J. Mathematical Capabilities of ChatGPT.  (2023)
\bibitem{sima2023arxiv}Sima, A. \& Farias, T. On the Potential of Artificial Intelligence Chatbots for Data Exploration of Federated Bioinformatics Knowledge Graphs.  (2023)
\bibitem{celerdata2023}Celerdata, B. ChatGPT is now finding bugs in databases. {\em  Https://celerdata.com/blog/chatgpt-is-now-finding-bugs-in-databases}. (2023)
\bibitem{borji2023arxiv}Borji, A. A Categorical Archive of ChatGPT Failures.  (2023)
\bibitem{iso2021mt}International Organization for Standardization ISO/IEC/IEEE 29119-4, Software testing - Part 4: Test techniques. {\em Software And Systems Engineering}. (2021)
\bibitem{ahlgren2021}Ahlgren, J., Berezin, M., Bojarczuk, K., Dulskyte, E., Dvortsova, I., George, J., Gucevska, N., Harman, M., Lomeli, M., Meijer, E., Sapora, S. \& Spahr-Summers, J. Testing Web Enabled Simulation at Scale Using Metamorphic Testing. {\em 2021 IEEE/ACM 43rd ICSE-SEIP}. pp. 140-149 (2021)
\bibitem{chen2017}Chen, T., Kuo, F., Liu, H., Poon, P., Towey, D., Tse, T. \& Zhou, Z. Metamorphic Testing: A review of challenges and opportunities. {\em ACM Computing Surveys}. \textbf{51} (2018)
\bibitem{martinez2022}Martinez-Fernandez, S., Bogner, J., Franch, X., Oriol, M., Siebert, J., Trendowicz, A., Vollmer, A. \& Wagner, S. Software Engineering for AI-Based Systems: A Survey. {\em ACM Transactions On Software Engineering And Methodology}. \textbf{31} (2022)
\bibitem{zhang2023chatgpt}Zhang, Y., Towey, D. \& Pike, M. Automated Metamorphic-Relation Generation with ChatGPT: An Experience Report. {\em 2023 IEEE 47th Annual Computers, Software, And Applications Conference (COMPSAC)}. pp. 1780-1785 (2023)



\end{thebibliography}


\end{document}